# Transient Analysis of Warm Electron Injection Programming of Double-Gate SONOS Memories by means of Full-Band Monte Carlo Simulation

G. Giusi[a], G. Iannaccone[b], U. Ravaioli[c]

[a]DEIS, University of Calabria, Via P. Bucci 41C, I-87036 Arcavacata di Rende (CS), Italy

[b]DIIEIT, University of Pisa, Via Caruso 16, I-56126 Pisa, Italy(g.iannaccone@iet.unipi.it)

[c]University of Illinois at Urbana-Champaign, Urbana, IL 61801 USA

ABSTRACT

In this paper we investigate "Warm Electron Injection" as a mechanism for NOR programming of double-gate SONOS memories through 2D full band Monte Carlo simulations. Warm electron injection is characterized by an applied $V_{DS}$ smaller than 3.15 V, so that electrons cannot easily accumulate a kinetic energy larger than the height of the Si/SiO$_2$ barrier. We perform a time-dependent simulation of the program operation where the local gate current density is computed with a continuum-based method and is adiabatically separated from the 2D full Monte Carlo simulation used for obtaining the electron distribution in the phase space. In this way we are able to compute the time evolution of the charge stored in the nitride and of the threshold voltages corresponding to forward and reverse bias. We show that warm electron injection is a viable option for NOR programming in order to reduce power supply, preserve reliability and CMOS logic level compatibility. In addition, it provides a well localized charge, offering interesting perspectives for multi-level and dual bit operation, even in devices with negligible short channel effects.

# I. INTRODUCTION

Multiple-gate and discrete-storage Non-Volatile Memories (NVM) offer the combined advantage of improved retention due to the suppression of Stress Induced Leakage Currents (SILCs), improved short channel effects, and reduced inter-cell capacitive couplings. These aspects make them particularly promising for aggressive downscaling into the nanoscale regime, and justify a significant research effort [1-4]. However, reliability concerns may limit the maximum longitudinal electric field and therefore the maximum applicable drain voltage $V_{DS}$ during NOR programming. Interestingly, experiments suggest that "warm electron injection", where electrons cannot accumulate a kinetic energy higher than the Si/SiO$_2$ barrier height ($V_{DS}$ < 3.15 V), is still much more efficient than Fowler-Nordheim programming and therefore can represent a reasonable option for NVM programming [4-6]. In a previous work [7] we studied the problem of warm electron injection in SONOS memories showing a strong dependence of the injected current on the drain voltage in the initial phase of the program operation, when the nitride layer is neutral. In this work we aim to investigate the time-dependent program operation in the warm electron injection regime and the dynamic trapping and detrapping of electrons in the silicon nitride layer. We propose a simulation methodology based on adiabatically decoupling the relatively slower process of charging the nitride layer from the faster process of electron transport in the MOSFET channel. Transport in nanoscale MOSFETs, in far from equilibrium conditions, can be accurately modeled by solving the Boltzmann Transport Equation (BTE) with the full band Monte Carlo device simulator MoCa [8]. By studying gate charge injection during the charging transient we are able to gather insights of the evolution of trapped charge. This information is particularly important for dual bit operation where physical charge localization is used to store more than 1 bit per cell. Because data information is stored in the oxide-nitride-oxide (ONO) stack through gate tunneling, accurate modeling of the gate current in extremely important for evaluating device performance. However, the gate current is several orders of magnitude smaller than the drain current, and its calculation poses a tremendous challenge to particle-based methods. Attempts to solve the BTE for the gate current problem were made [9-10]. An energy transport model and a Monte Carlo approach were successfully applied to gate current calculations for the case of hot carrier injection [11-14].

Approaches to the simulation of the time evolution of the charge injected into the nitride are found in the literature. For example, in Ref [15] only hot electrons are considered and all the injected charge is considered trapped. In Ref. [16] a trapping-detrapping model includes thermal excitation as the main discharge mechanism. Reference [17] proposes a method to accelerate the iterative Monte Carlo procedure. In Ref. [18] the stored charge is evaluated as the difference between the injected charge and the charge emitted via the Poole-Frenkel effect. In this work we use the Monte Carlo approach to calculate charge density and potential, and the transmission coefficient for each point of the silicon/silicon oxide interface is calculated using the WKB approximation. Quantum confinement in the channel and barrier lowering are not considered.

Differently from the cases mentioned above, we calculate gate injection contributions also for "warm" carriers, whose kinetic energy is lower than the barrier height. Charge trapping/detrapping in the nitride layer is taken into account by a Shockley-Read-Hall model where the trap-to-band tunneling is considered as the main discharge mechanism.

Simulation results show that injection is effective also for low drain bias due to the very strong dependence of the gate current on $V_{DS}$. Charge injection is well localized, offering interesting perspectives for dual bit and multi bit operation even in devices with reduced short channel effects (SCE) such as multi-gate devices. Warm electron injection could be very useful for increasing reliability, reducing supply voltage and hence power dissipation.

The remainder of the paper is organized as follows. In Section II the physical model of gate tunneling and of trapping/detrapping in the nitride layer is described. In Section III the simulation methodology to calculate the time and space dependent gate current density and stored charge density is presented. An application of the method on double-gate SONOS memories is shown in Section IV. Conclusions are drawn in Section V.

## II. PHYSICAL MODEL

In this Section the adopted physical model is presented. In Section IIA we discuss tunneling of electrons from the channel through the Si/SiO2 interface, while in Section IIB we discuss the physical model of charge trapping-detrapping in the silicon nitride layer.

## IIA. PHYSICAL MODEL OF ELECTRON TUNNELING

As stated in the previous Section, we consider not only "hot electrons" with total kinetic energy higher than the Si/SiO$_2$ interface barrier height $B = 3.15$ eV, but also warm electrons with lower kinetic energy that provide the major contribution to the tunneling current for low $V_{DS}$ programming. As in [7], and differently from previous contributions in the literature, we need to compute the local tunneling current at each point of the Si/SiO$_2$ interface, because injection is highly non uniform in space and because charge trapping is localized. Our tunneling model is relatively simple: we assume that i) total energy and transverse momentum are conserved during tunneling; ii) the dispersion relation in the SiO$_2$ is parabolic with isotropic effective mass m$_{ox}$; iii) the transmission coefficient can be computed with the WKB approximation.

The considered structure is sketched in Fig. 1a and is the same of Ref. [7]. In the diagram, $x$ is the channel direction and $y$ is the direction of tunneling (perpendicular to the Si/SiO$_2$ interface). In the following with refer to the interface positioned at $y=0$, with $y>0$ for silicon and $y<0$ for oxide. $E$ is the total carrier energy and ($k_x$, $k_z$) is the transverse wave vector. If $E_{kin}=E_{kin}(0^+)$ is the kinetic energy of the particle at the interface on the silicon side, then the kinetic energy of the particle at the interface on the oxide side $E_{ox\_kin}= E_{kin} (0^-)$ is:

$$E_{ox\_kin} = E_{kin} - B \qquad (1)$$

The dispersion relation in the oxide is assumed to be:

$$E_{ox\_kin} = \frac{\hbar^2}{2m_{ox}}\left(k^2_{ox\_x} + k^2_{ox\_y} + k^2_{ox\_z}\right) \qquad (2)$$

where $k_{ox\_x}$, $k_{ox\_y}$, $k_{ox\_z}$ are the components of wave vector. We should note that in the case of tunneling $E_{ox\_kin}$ is negative and $k_{ox\_y}$ is purely imaginary, while $k_{ox\_x}$, $k_{ox\_z}$ are real because they are conserved during tunneling ($k_{ox\_x} = k_x$, $k_{ox\_z} = k_z$). From (2) we obtain $k_{ox\_y}$, and the component of the kinetic energy in the oxide along the tunneling direction is:

$$E_{ox\_y} = \frac{\hbar^2}{2m_{ox}} k_{ox\_y}^2 = E_{ox\_kin} - \frac{\hbar^2}{2m_{ox}} \left(k_x^2 + k_z^2\right) \tag{3}$$

The component of the kinetic energy contributing to tunneling, $E_y$, can be separated as

$$E_y = E_{kin} - \frac{\hbar^2}{2m_{ox}}\left(k_x^2 + k_z^2\right), \tag{4}$$

so that the effective barrier height is identified as $\phi_s = B - E_y$.

We consider for tunneling only particles that are at the Si/SiO$_2$ interface and have a positive velocity in the tunneling direction ($v_y$). During the Monte Carlo solution of the BTE, for each particle, we can calculate $E_y$ from (4) and compute the distribution $n(x,E_y)$, where $n(x,E_y)dE_y$ is the density per unit volume of electrons in $y=0$ that have component of the kinetic energy between $E_y$ and $E_y + dE_y$. The tunneling current density can be calculated using the formula

$$J_G(x) = q \int v_y n(x, E_y) T(E_y) dE_y. \tag{5}$$

References [19] and [20] have shown that with proper barrier parameters the I-V characteristics of thin gate stacks can be reproduced with reasonable accuracy of several orders of magnitude without taking into account barrier lowering and with the WKB approximation. We therefore compute the transmission coefficient $T=T(E_y)$ as in Ref [19].

## IIB. PHYSICAL MODEL OF TRAPPING DETRAPPING

NVM SONOS memories operate by storing charge in localized states in the nitride of the ONO stack. At the moment no clear consensus exists about nitride trap distribution in energy and space. As the reader can imagine this information is very important to model the stored charge which in turn influences the threshold voltage shift and thus the "stored information". Typically, a trap profile is determined by retention experiments and by simulation fitting [21-26], and consists in a uniform trap distribution in space and a constant trap energy comprised between 0.8 eV and 1.4 eV below the nitride conduction band (CB). Here, we make the same assumption considering a mono-energetic level $E_t$ =1.0 eV below the nitride conduction band as suggested in Ref. [21]. Generation-Recombination in the nitride is governed by the Shockley-Read-Hall (SRH) generation-recombination [27]. Generally, traps can be filled or emptied by electron/hole capture/emission processes. We neglect hole contribution because a) in the gate there are few holes that can tunnel into the nitride, b) the valence band shift (4.1eV) is higher than the conduction band shift (3.15 eV) at the Si/SiO$_2$ interface.

The processes considered here are illustrated by Fig. 2. Traps in the nitride can be filled by channel electrons which tunnel through the Si/SiO$_2$ barrier. These electrons, depending on their kinetic energy $E_y$ along the tunnel direction at the interface, can be trapped directly (process B in the figure) or indirectly by thermal recombination (process A+C). Detrapping is due to two processes [21]: 1) thermal emission due to electrons which emit toward the conduction band (process D), 2) trap-to-band tunneling due to electrons trapped which tunnel directly into the gate conduction band (E). Other charge loss processes as band-to-trap tunneling, trap-to-trap tunneling and Poole-Frenkel emission have been neglected [21-22]. We neglect also the redistribution of charge between nitride traps which was found to be governed by Poole-Frenkel conduction [28], because this process is too slow with respect to the processes we have considered to be relevant during programming [21,28-30]. Thermal recombination (A+C) is found to be dominant with respect to process B, due to the much higher transmission coefficient, whereas process E is dominant with respect to processes D+F, due to the very low thermal generation rate at room temperature [21]. In summary, processes B and D+F have been neglected and we consider only A+C and E.

The nitride region is subdivided into spatial bins along the *x* and *y* directions. For each *x,y* bin the SRH equation is:

$$\frac{dn_T(x,y)}{dt} = c(x)p_T(x,y) - e(x,y)n_T(x,y) \tag{6}$$

where $n_T$ is the concentration of occupied traps, $p_T$ is the concentration of free traps, $c(x)$ is the capture rate and $e(x,y)$ is the emission rate. The capture rate is calculated from the gate current density given by (5):

$$c(x) = \frac{\sigma}{q} J_G(x) \tag{7}$$

where σ is the capture cross-section. The emission rate $e(x,y)$ is given by $e(x,y) = e_{TBT0} T_G(x,y)$, where $e_{TBT0}$ is the "attempt-to-escape frequency" and $T_G(x,y)$ is the transmission probability (calculated through the WKB approximation) through the control barrier between the considered trap and the gate CB.

A complete charge trapping-detrapping model should include also a transport model in the nitride. This is a very complicated task and many parameters like the exact trap density $N_T$, the trap cross section, energy relaxation time for electrons, electron mobility in silicon nitride have to be known accurately. As stated before, several different values can be found in the literature for those parameters, that in the end are all extracted through fitting with experiments: typical extracted values for $N_T$ are in the range $5 \times 10^{18}$-$10^{19}$ cm$^{-3}$, and for σ are in the range $5 \times 10^{-12}$-$5 \times 10^{-13}$ cm$^2$ [21-26].

We believe that in our case we can make a reasonable simplification to strongly reduce the number of free parameters, assuming that all injected charge is trapped uniformly throughout the silicon nitride layer. Our assumption would not hold in general but it is based on the fact that in our case electrons are injected into the nitride layer with a relatively low kinetic energy, and can lose much of this energy before reaching the high control oxide barrier and being reflected by it. The assumption of uniform trapping of injected electron in the layer thickness is acceptable for a very thin layer as in our case, and it allows us not to consider transport in the nitride layer explicitly. This means that $\sigma N_T t_N = 1$ and that the SRH equation rate (6) reduces to:

$$\frac{dn_T(x,y)}{dt} = \frac{J_G(x)}{qt_N} - e_{TBT0} T_G(x,y) n_T(x,y) \tag{8}$$

where $N_T$ is the trap density and $t_N$ is the nitride thickness. In our trapping-detrapping model now only two free parameters remain, $E_t$ and $e_{TBT0}$, which determine the behavior of trap-to-band tunneling process (8).

In the pioneering work of Lundkvist [31] on charge loss in SONOS, the attempt-to-escape frequency was expressed as $e_{TBT0} = E_t / h$ and the transmission coefficient $T_G$ was calculated by assuming a rectangular barrier. The former assumption would have a physical basis only if the trap energy considered was taken with respect to the bottom of the conduction band in a potential well, and not with respect to the conduction band out of the well, as in this case (For $E_t$=1eV one would obtain $e_{TBT0}$=2.4·10$^{14}$ s$^{-1}$ ). Several values for $e_{TBT0}$ can be found in literature in the range 10$^{12}$-10$^{14}$ s$^{-1}$ [21, 24-26, 31]. Here we prefer to use the value extracted in Ref. [25] from the comparison between retention experiments and simulations ($e_{TBT0}$= 2·10$^{12}$ s$^{-1}$).

The assumption of rectangular barrier in the control oxide, is very commonly made [21, 23-26] but is not very realistic, especially in modern devices, where the oxide field can be close to 1 V/nm. For this reason, we compute in detail $T_G(x,y)$ with the WKB approximation.

## III. SIMULATION METHOD

Device simulation was performed with the 2D full band Monte Carlo (MC) simulator "MoCa" [8]. MoCa includes all relevant scattering mechanisms and it has been purposely modified to include the simulation of the gate current with a continuum-based method (as opposed to the particle-based method used to compute particle distributions and transport in the channel): the gate current was computed by extracting particle energy distribution and potential by MoCa and substituting it in (5).

The problem of calculating the charge trapped in the nitride can be addressed by solving the time dependent SRH equation rate (8), where capture and emission rates depend on the occupation density $n_T$, which changes with the time. In order to perform a time dependent simulation, we adiabatically decouple transport in the channel from electron tunneling into and from the nitride, because the former is a faster mechanism. In addition, we assume that the tunneling current is a negligible fraction of the drain current, and do not consider it explicitly when computing transport properties in the channel.

At each timestep, electron transport is computed by MoCa in DC conditions assuming that the charge in the nitride layer is fixed. The electron distribution obtained at the interface and the potential profile are used to calculate the new capture and emission rates (Fig. 1b). Now, the new occupancy $n_T$ can be calculated by forward integration of (8) with an appropriate choice of an adaptive timestep $\Delta t$. Finally, the total charge in the nitride layer is updated and the simulation continue with the next MC timestep until the desired programming time is reached.

An important issue can arise when evaluating the particle distribution with a 2D MC simulator regarding particle-particle interaction, an issue discussed already in our previous work [7]. Here we would like to just summarize our conclusions by saying that we are aware of the fact that there are accurate approaches to naturally include short range particle-particle interaction in Monte Carlo simulation with a 3D solver, as implemented for instance in the simulator developed by one of the authors of this work [32]. The cost of 3D solutions remains, however, prohibitive and in order to reduce the computational complexity of the problem we had to limit our approach to the 2D full-band version of MoCa, where we approximately take into account particle-particle interaction by self-consistently solving the Poisson equation on a relatively fine grid, without explicitly introducing electron-electron scattering mechanisms [33].

IV. SIMULATION RESULTS

To validate the proposed simulation method and the physical model, we used as test structure an n-channel DG SONOS memory with a 50 nm channel length and a 4/5/5 nm ONO stack. The acceptor Fin Doping is $3 \cdot 10^{18}$ cm$^{-3}$, while the source/drain doping extends under the gate for 15 nm on each side (Fig. 1a).

In Fig. 3 we show the distribution of the kinetic energy $E_{kin}$ for electrons at the silicon-oxide interface in various positions along the channel at time t = 0 s and for a bias of $V_{GS}$=8V and $V_{DS}$=2.8V. As can be seen, only at the source (x=0 nm) electrons obey a Maxwell-Boltzmann distribution. For x>0 nm, the distribution is progressively more asymmetric, with a flatter profile for energies smaller than the potential energy drop with respect to the source, and with un upper thermal equilibrium tail for larger kinetic energies shaped according to the lattice temperature. Since transport is partially ballistic, for a drain voltage $V_{DS}$, a significant portion of electrons injected from the source have in the vicinity of the drain a kinetic energy $qV_{DS}$ so that they see a barrier towards the gate of approximate height $B - qV_{DS}$. Such effective barrier lowering significantly increases local tunneling close to the drain.

In Fig. 4 we plot local capture and emission rates at $t = 0$ s for a fixed $V_{GS}$=8V and for different $V_{DS}$. The capture rate has been calculated by assuming the capture cross section σ to be $5 \cdot 10^{-13}$ cm$^2$ [22] and the emission rate (which depends on $y$) has been computed at the interface between nitride and the control oxide. It is apparent that rates are very sensitive to $V_{DS}$ and that carrier injection is localized at the drain side. Also, only at this side carrier trapping is possible because capture rate dominates over emission rate. Farther away from the drain (roughly about 10 nm) emission dominates over capture and charge trapping will not occur. This might suggest the possibility of using drain voltage for multi-level programming with $V_{DS}$<3.15 V.

As time progresses, capture and emission processes become comparable at the drain side and charge injection saturates, as can be seen in the plot of the tunneling currents through the two oxides as a function of the programming time (Fig. 5). As can been seen in Fig. 6, the maximum of the injected current density, for a fixed $V_{GS}$=8V, decreases exponentially with time, with a behavior independent of $V_{DS}$. During programming, the trapped charge (Fig. 7) remains rather localized on the drain side. We can define the effective size of the charge storage region as the ratio of the total stored charge to the maximum stored charge density. This quantity is plotted in Fig. 8: as one can see, it does not change significantly during programming operation but it increases slightly for higher $V_{DS}$ values. Threshold voltage shifts in forward and reverse read operation (Fig. 9) are sensitive to $V_{DS}$ confirming that low $V_{DS}$ programming can be used in place of conventional CHEI for dual bit operation as also confirmed by experiments [6]. Moreover, Fig. 10 emphasizes that the programming time for a given required $V_T$ shift is reduced by up to two orders of magnitude for an increase of 1 V of $V_{GS}$, and for the same programming time the $V_T$ shift increases by 0.3-0.4 V for each Volt of increase of $V_{GS}$. Also, this result is in agreement with [6]. Let us stress the fact that the comparison between our simulations and Ref. [6] can only be qualitative because we are considering double-gate devices while experiments in [6] regard trigate devices where injection at the corners (especially for high fields) can be largely dominant [34].

## IV. CONCLUSIONS

We have developed a simulation methodology based on a Monte Carlo device simulator to investigate the "warm electron injection" programming regime ($V_{DS} < 3.1$ V) of NOR DG SONOS based on the adiabatic separation of electron transport in the channel with respect to trapping/detrapping in the ONO layer. The gate current is calculated as a post-processing step of the Monte Carlo simulation by a continuum based method in conjunction with the particle-based method used to compute particle distributions and transport in the channel. We have shown that warm electron injection is potentially more effective than FN NAND programming, and that stored charge is well localized, providing a significant forward-reverse threshold voltage window both for multi bit and for multi level operation. Warm electron injection should be a viable option for NOR programming, preserving reliability, power dissipation and CMOS logic level compatibility, at the cost of a slower programming time with respect to hot electron operation.

## ACKNOWLEDGMENT

This work was supported by the FinFlash IST-NMP Project, the FIRB Project RBIP06YSJJ, and by the RHESSA Project of the Italian Ministry for Foreign Affairs.

FIGURE CAPTIONS

Figure 1a

The simulated structure, an n-channel DG SONOS memory with a 50 nm channel length and a 4/5/5 nm ONO stack. The acceptor Fin Doping is $3 \cdot 10^{18}$ cm$^{-3}$, while the source/drain doping extends under the gate for 15 nm from each side. $x$ is the channel direction and $y$ is the direction of tunneling (perpendicular to the Si/SiO$_2$ interface). One interface is at $y=0$, $y>0$ is silicon, $y<0$ is oxide.

Figure 1b

Simulation flowchart. Electron energy distribution and potential are extracted by MC simulation and used to calcolate the injected charge and capture and emission coefficients. The SRH equation rate (Eq. 8) is solved with a convenient choice of an adaptive time step $\Delta t$. Next the total charge into the nitride is updated with the trapped charge obtained from Eq. 8. The cycle restarts with the next MC simulation until the desired programming time is reached.

Figure 2

Energy band diagram and mechanisms involved during program operation.: A) Electron injection from the Silicon CB to the Nitride CB, B) Tunneling and capture from the Silicon CB to nitride traps, C) Thermal Recombination from the nitride CB to the nitride traps, D) Thermal Emission from the nitride traps to the nitride CB, E) Trap to band tunneling from the nitride traps to the Gate CB, F) Electron injection from the nitride CB to the Gate CB. Only A+C and E have been considered in our simulation.

Figure 3

The distribution of the kinetic energy $E_y$ at the beginning of the injection ($t$=0s) for electrons at the silicon-oxide interface in various positions along the channel and for a fixed bias. Only at the source side the distribution is a displaced maxwellian.

Figure 4

Capture and emission rates at the top oxide/nitride interface along the channel and for different $V_{DS}$ values (0V, 1V, 2V, 3V, 4V) at the beginning of the injection (t=0s). Charge trapping is possible only on the drain side where capture rate is several orders of magnitudes higher than the emission rate.

Figure 5

Current density along the channel through the tunnel oxide (in) and through the control oxide (out) for increasing time during the program operation.

Figure 6

The max of $J_G$ as function of the programming time is quite independent of the drain bias and has roughly a 1/t behavior in a log-log scale.

Figure 7

Trapped charge density in the nitride layer along the channel as function of the programming time.

Figure 8

Effective injection length as a function programming time for different values of $V_{DS}$. As time proceeds no significance change in injection length occurs.

Figure 9

Threshold voltage displacement in forward and reverse read as function of the programming time for a fixed $V_{GS}$=8 V and for different $V_{DS}$. It is evident that dual bit operation is possible also with $V_{DS}$<3.15 V.

Figure 10

Threshold voltage displacement in forward and reverse read as function of the programming time for a fixed $V_{DS}$=2.8 V and for different $V_{GS}$. Multi-level operation seems to have larger window respect to multi-bit operation

**FIGURES**

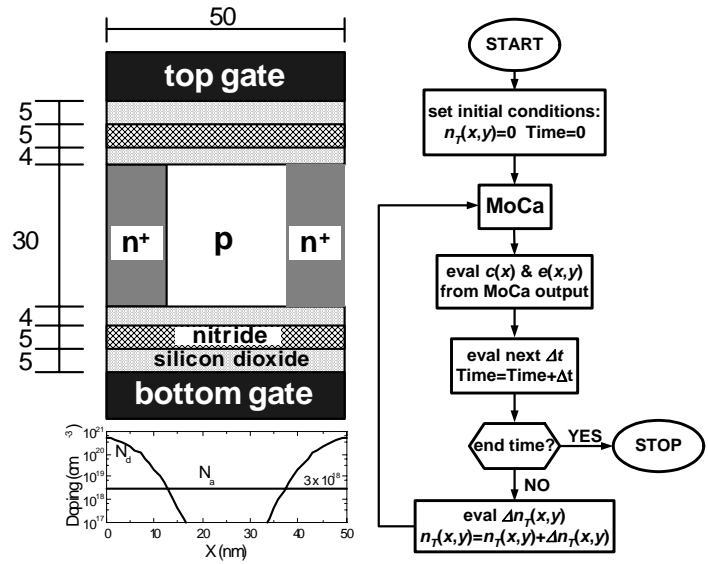

Figure 1a, b

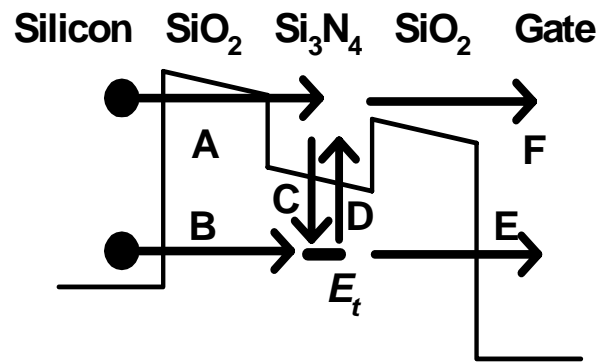

Figure 2

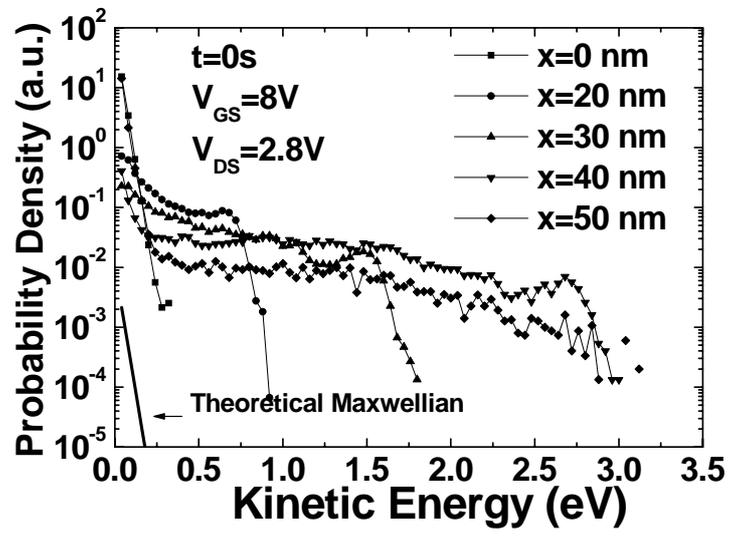

Figure 3

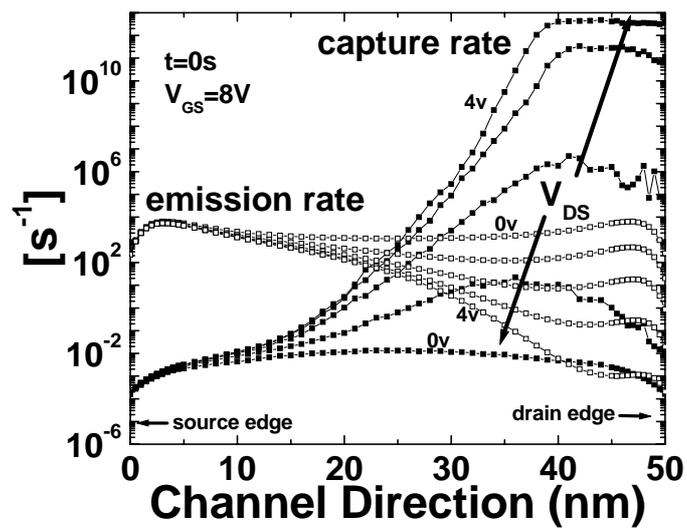

Figure 4

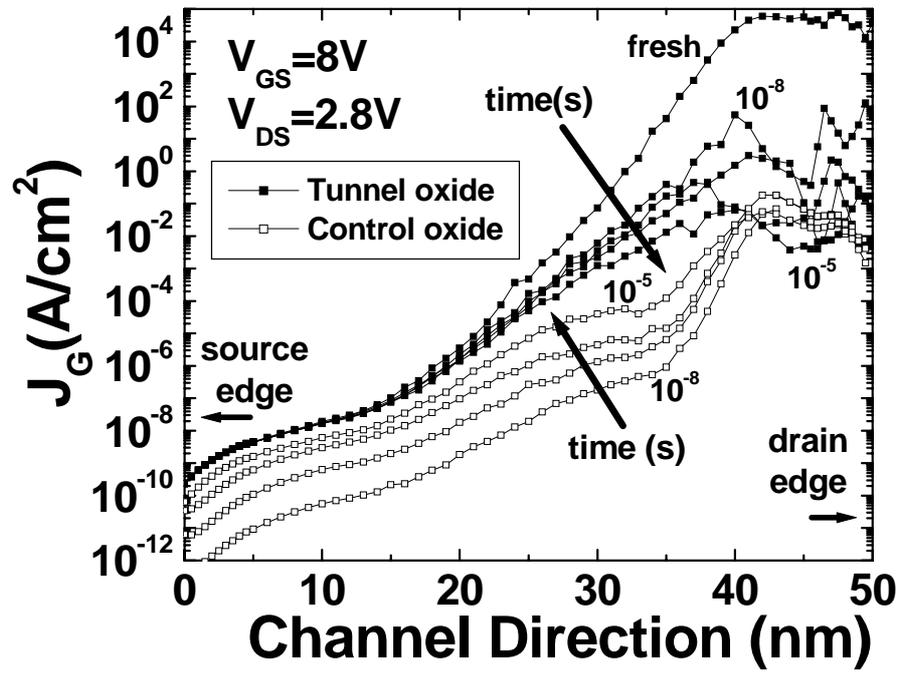

Figure 5

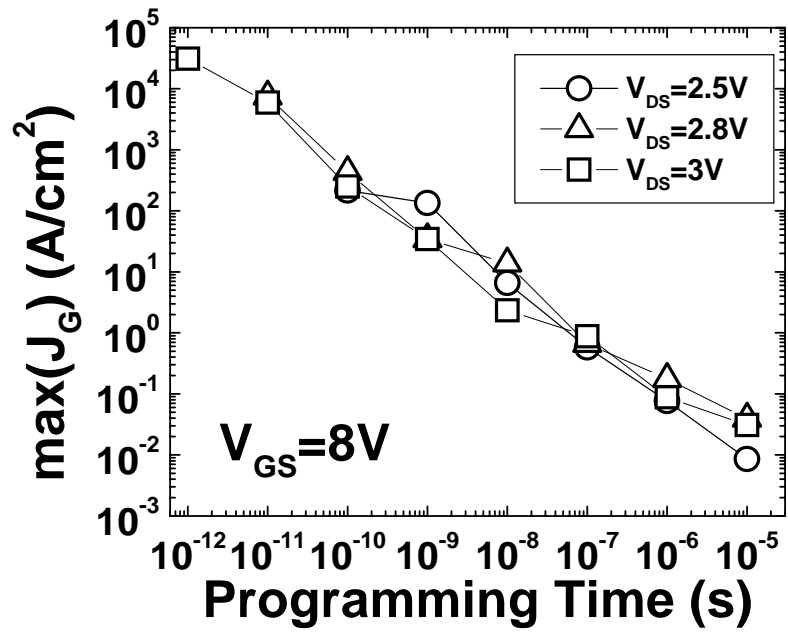

Figure 6

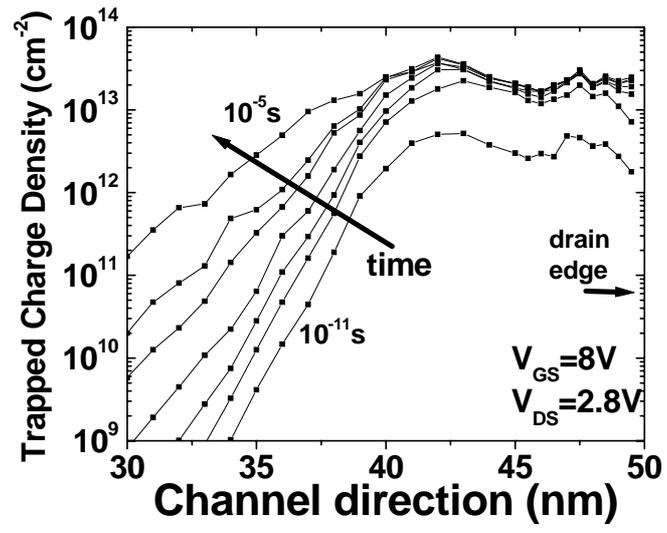

Figure 7

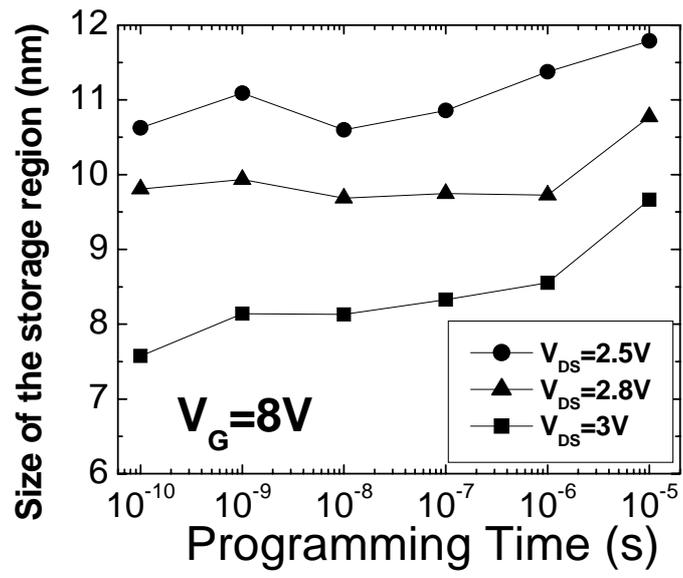

Figure 8

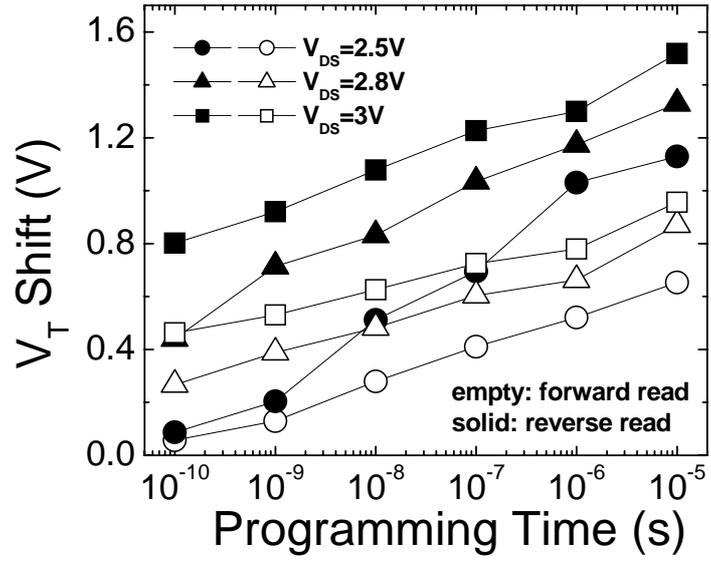

Figure 9

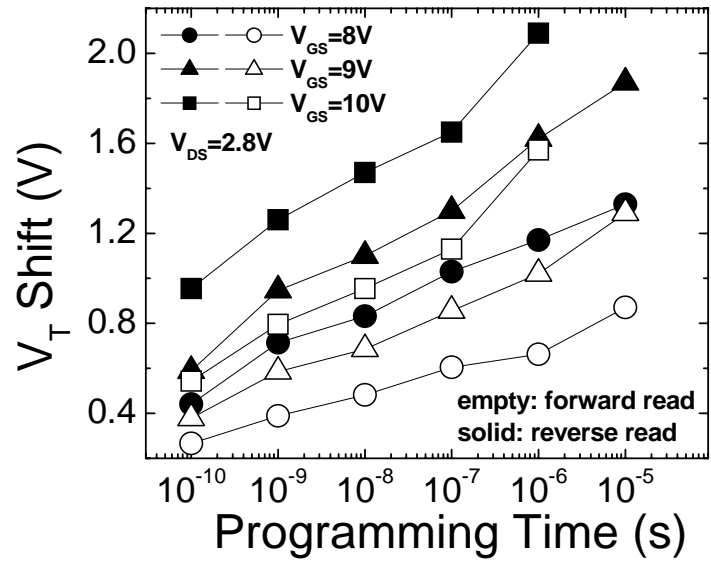

Figure 10